# Organisational accounts engaged in scholarly communication on Twitter: Patterns of presence, activity and engagement

Zohreh Zahedi[1,3], Yanqing Zhang[2], Zekun Han[2], Er-Te Zheng[4], Zhichao Fang[2,3]*

* Corresponding author: z.fang@cwts.leidenuniv.nl

1. Department of Information Science, Faculty of Humanities, Persian Gulf University, Bushehr, Iran.

2. School of Information Resource Management, Renmin University of China, Beijing, China.

3. Centre for Science and Technology Studies (CWTS), Leiden University, Leiden, the Netherlands.

4. School of Information, Journalism and Communication, the University of Sheffield, Sheffield, UK.

**Abstract**: Organisational accounts are an integral part of the Twitter (now X) ecosystem. This study identified 9,842 research- and policy-related organisational accounts that had tweeted about scholarly publications by linking three global organisational databases (GRID, ROR, and Overton) with two altmetric databases containing Twitter data (Altmetric and the former Crossref Event Data). The resulting openly available dataset was used to examine organisational activity in scholarly communication across three dimensions: social media capital, tweeting activity, and engagement level. The results show that, compared to all Twitter users engaged in scholarly communication, organisational accounts hold a notable advantage in terms of follower bases and the proportion of scholarly tweets. Their scholarly tweets achieve high visibility through likes and retweets but perform weakly in generating more conversational forms of engagement, such as quotes and replies. Distinct patterns emerge across organisational categories: research facilities, in particular, demonstrate the strongest focus on scholarly tweeting, whereas government accounts are comparatively more successful in eliciting engagement across all metrics, including the more interactive ones. This study contributes both an open dataset of organisational accounts and a methodological framework for their identification, while also highlighting the important roles that organisations play in shaping scholarly discourse on social media.

## 1. Introduction

Social media has increasingly played an important role in informal scholarly communication and in fostering public engagement with science [1]. These developments create opportunities to trace broader forms of attention to scholarly outputs beyond the academic community. As a result, the field of social media metrics has emerged, focusing on the measurement of online activity surrounding scholarly entities – such as scholarly publications and researchers – across a variety of platforms [2]. The objectives of these metrics include enhancing our understanding of public engagement with science and exploring their potential applications in research evaluation, scholarly communication, and broader societal engagement [3].

Unlike traditional academic citations, which are primarily generated by members of the academic community who possess the expertise and motivation to produce research outputs, social media activity surrounding scholarly work can originate from a wide variety of actors. These range from academics and students to journalists and members of the general public [4–6]. This heterogeneity presents both opportunities and challenges for social media metrics [7]. On the one hand, it enables the capture of attention and engagement from broader societal stakeholders. On the other hand, it complicates the interpretation of social media attention, since different types of users may be driven by distinct motivations when engaging with scholarly content.

To address the interpretive challenges posed by this heterogeneity, numerous studies have sought to identify and classify the types of social media users involved in the dissemination of scholarly outputs. Within the diverse and continuously evolving social media landscape, this line of inquiry is particularly important for deepening our understanding of the nature of online attention to scholarly outputs and for assessing the implications of social media metrics in the context of expanding research evaluation frameworks. Existing studies have generally pursued two main approaches: (1) classifying different types of users directly from social media datasets, and (2) identifying particular categories of users based on pre-compiled lists of entities.

*1.1. Classification of different user types*

Within the first research workflow, many studies have classified social media users engaged in scholarly communication into different typologies. Three primary methods have been employed: manual annotation, content analysis, and network-based approaches.

Manual annotation is typically applied to relatively small datasets. In this approach, researchers classify users according to codebooks and self-developed schemes, often drawing on information from users' social media profiles [5, 6, 8–13]. While manual annotation provides detailed and context-sensitive categorization, it is time-consuming and therefore usually limited to small samples. For larger datasets, researchers have generally relied on automated or semi-automated methods, such as content analysis of user biographies or the construction of social networks based on user interactions.

User biographies (bios) provide direct evidence of self-presentation across dimensions such as occupation, academic title, gender, and personal or professional interests. Accordingly, many studies have classified users into identity groups by analyzing the textual content of bios. Approaches include keyword matching [14], machine learning [15], co-word analysis [16–18], and topic modelling [19]. A notable example is the classification system used by Altmetric, which categorizes Twitter users into four broad types: researchers, practitioners, science communicators, and members of the public [20, 21].

Beyond textual analysis, social networks among users offer a more indirect means of classification. Because users with similar backgrounds or interests tend to interact more frequently, relationships such as following, mentioning, and retweeting can be used to construct networks and identify communities of users who share disciplinary affiliations or topical interests [22–28].

Collectively, these studies underscore the diversity of participants involved in the social media dissemination of scholarly outputs. Both academic actors (e.g., researchers, journals, and scholarly societies) and non-academic users (e.g., members of the public) actively share, discuss, and circulate scholarly work [5, 6, 18]. Importantly, these groups are often motivated by distinct factors [4, 29–31], reinforcing the idea that social media engagement with scholarly content is complex and must be interpreted with careful attention to user demographics.

Nevertheless, this first research workflow has notable limitations. Manual annotation is restricted to small samples, while clustering approaches based on user bios or social networks often yield only coarse-grained categories. As a result, these methods may struggle to assign users to clearly defined identities or to capture the nuanced motivations that underlie their engagement with scholarly content.

*1.2. Identification of specific types of users*

In contrast to the classification approach, the second research workflow aims to generate more explicit datasets of particular types of users. Studies in this area often begin with a predefined list of entities – such as scholars, scholarly journals, or universities – and then identify which of these entities are active in scholarly communication on social media [32]. A typical example is the identification of scholars who use Twitter to share scholarly publications. For instance, previous research has matched the bibliographic information of authors with publications indexed in the *Web of Science* or *OpenAlex* against the demographic and profile information of Twitter users sharing scholarly content, thereby constructing dataset of scholars active on Twitter [33, 34]. Beyond author name matching, some studies have also used occupation-related terms such as "biologist" or "economist" to retrieve Twitter lists named after these occupations, which can then be used to identify potential scholar users included in those lists [35, 36].

In addition to individual academics, several studies have examined the presence of scholarly journals on social media. Researchers have curated lists of journal names and searched for them across platforms such as Twitter, Facebook, and WeChat to assess the uptake and activity of journals. Reported adoption rates vary depending on the platform and journal set studied, ranging from 7.2% of journals in *Social Science Citation Index* (SSCI) with a Facebook presence [37], to 22.2% of journals in *Science Citation Index-Expanded* (SCIE) with Twitter accounts [38], and as high as 84.4% of journals in *Chinese Science Citation Database* (CSCD) with WeChat accounts [39]. Beyond journals, researchers have also investigated the social media presence of other organisational types, including universities and departments [40–42], academic libraries [43, 44], scholarly book publishers [45], and companies [46]. These studies provide valuable insights into how different organisations adopt social media tools to increase their visibility.

Another area of particular focus has been bot accounts, which play an important role in the online dissemination of scholarly content [47]. In addition to manual identification, automated tools such as *Bot or Not?* and *BotometerLite* have been applied in identifying bots in the context of scholarly communication [48, 49]. Although bots account for only a small fraction of users tweeting about science (estimated between 0.2% to 1.2%), they contribute a disproportionately large share of tweets containing scholarly content (ranging from 2.9% to 51.3%) [6, 14, 26, 48–52]. This amplification is especially pronounced in fields such as the life and earth sciences, where bots substantially boost the volume of scholarly tweets [9].

*1.3. Objectives of the study*

Previous research has made substantial contributions to understanding the social media presence of various user groups and the roles they play in online scholarly communication. However, comparatively little attention has been given to organisations, despite their importance as institutional actors linking research production, policy processes, and public engagement. As noted earlier, open datasets have been developed for several user types, including individual researchers, scholarly journals, and bots, which have greatly facilitated investigations into their tweeting behaviors and influence. Yet, a dedicated dataset of organisational Twitter accounts remains unavailable. This absence limits empirical research on an important class of actors whose activities may differ systematically from those of individuals.

Addressing this gap is valuable for several reasons. First, an open dataset of organisational accounts would enable future research to more effectively investigate how institutions communicate and shape scholarly discourse online, supporting work in scientometrics, altmetrics, and science communication. Second, such a dataset would benefit policymakers and research institutions seeking to understand how organisations contribute to the dissemination and visibility of scientific knowledge on social media. Third, a descriptive analysis of organisational tweeting activity and engagement provides insight into the institutional dimension of online scholarly communication, complementing existing research that has largely focused on individual-level participation.

To achieve these goals, the present study focuses on Twitter[1] and examines how

---

[1] It is important to note that Twitter was the official name of X before its acquisition by Elon Musk. As

research- and policy-related organisations use the platform to engage in online communication surrounding scholarly publications. Specifically, we (1) construct an open dataset of Twitter accounts operated by organisations that have tweeted about scholarly publications, and (2) address the following research questions (RQs):

- **RQ1**. How do the scholarly tweeting patterns of organisations compare with those of all Twitter accounts engaged in scholarly communication?

- **RQ2**. How do scholarly tweeting patterns vary across different categories of organisations?

**2. Data and methods**

To construct a dataset of organisational accounts involved in scholarly communication on Twitter, we performed a two-step process. First, we compiled a comprehensive list of organisations based on three major databases: *GRID*, *ROR*, and *Overton*. Second, we matched this list against Twitter accounts that had tweeted about scholarly publications, as recorded by two altmetric data aggregators: *Altmetric* and *Crossref Event Data*. This section describes how the organisational and Twitter account datasets were retrieved and how the matching process was carried out.

*2.1. Dataset of organisations*

To obtain a global dataset of research- and policy-related organisations, we integrated information from three complementary databases: GRID, ROR, and Overton. The following provides a brief introduction and the snapshot date of the version used for data retrieval for each organisational database:

- GRID (Global Research Identifier Database, https://www.grid.ac) is an open database of unique identifiers for research-related organisations, released by *Digital Science* in 2015. Although GRID was discontinued in 2021 in favor of its successor – the Research Organisation Registry (ROR) – its historical releases remain openly available via Figshare. From the GRID database maintained at the Centre for Science and Technology Studies (CWTS), Leiden University (snapshot: September 2021), we extracted 106,149 distinct organisations.

---

Twitter was the official name during our data collection in March 2023 for this study, we continue to use the term throughout the study, as well as the related terms such as tweet, retweet, like, quote, and reply.

- ROR (Research Organisation Registry, https://ror.org/) is an open database of persistent identifiers for research organisations, initially seeded from GRID but independently maintained since 2022. Operated collaboratively by the California Digital Library, Crossref, and DataCite, ROR provides identifiers and metadata for a wide range of global organisations. From the CWTS in-house ROR database (snapshot: August 2024), we extracted 138,546 distinct organisations.

- Overton (https://www.overton.io/) is a bibliographic database of policy documents launched in 2019. Overton indexes documents from policy-related organisations worldwide, including government agencies, intergovernmental organisations, and think tanks [53]. From the CWTS in-house Overton database (snapshot: July 2024), we retrieved 2,244 distinct organisations.

After cleaning and deduplication across the three sources, we compiled a final list of 111,280 unique organisations. For each organisation, we retained detailed metadata – such as full name, aliases, address, and website URL – where available from the source databases.

*2.2. Dataset of Twitter accounts*

To examine which organisations engage in scholarly communication on Twitter, we constructed a comprehensive database of *scholarly tweets*, defined as tweets that reference scholarly publications. Tweet IDs were sourced from two major altmetric data providers: Altmetric and Crossref Event Data. Below, we provide a brief overview of each data source, along with the snapshot date of the version used for data retrieval:

- Altmetric (https://www.altmetric.com), founded in 2011 and now part of *Digital Science*, is one of the most widely used altmetric data aggregators. Altmetric monitors a broad range of online sources to capture events related to research outputs, including news outlets, blogs, Mendeley, and social media platforms such as Twitter and Facebook. From the CWTS in-house Altmetric database (snapshot: November 2022), which was constructed using dump files provided by Altmetric, we retrieved 175,852,041 unique tweet IDs that referenced scholarly publications.

- Crossref Event Data (CED, https://www.crossref.org/services/event-data/) is an altmetric service maintained by Crossref and accessible via a public API. CED tracks online events surrounding research outputs from diverse sources including news outlets, blogs, Wikipedia, and social media platforms. Twitter data was included until February 2023. From the CWTS in-house CED database (snapshot:

January 2023), constructed using the CED API, we retrieved 75,286,156 tweet IDs of scholarly tweets.

In March 2023, we used the Twitter API to collect detailed metadata for all tweets recorded by both Altmetric and CED. After performing data cleaning and deduplication – including the removal of unavailable tweets, primarily due to deletions or account suspensions [54] – we compiled a final database of 150,586,368 scholarly tweets posted by 10,119,331 distinct Twitter accounts.

The resulting database contains detailed information on both tweets and accounts. For tweets, we collected engagement metrics such as the number of likes, retweets, quotes, and replies. For accounts, we retrieved profile information including user's name, profile website URL, self-reported geolocation, and counts of followers and followings.

*2.3. Matching process to identify organisational Twitter accounts*

Figure 1 presents the workflow used to identify organisational accounts that tweeted about scholarly publications. We first matched organisations and Twitter accounts based on organisational full names, aliases, and profile information. For each candidate pair of organisation and Twitter account, we conducted manual verification using additional information such as the organisation's address, the self-reported geolocation of the Twitter account, and Twitter links embedded in the organisation's official website.

Following this manual validation process, we identified a total of 9,842 Twitter accounts operated by organisations included in our datasets.

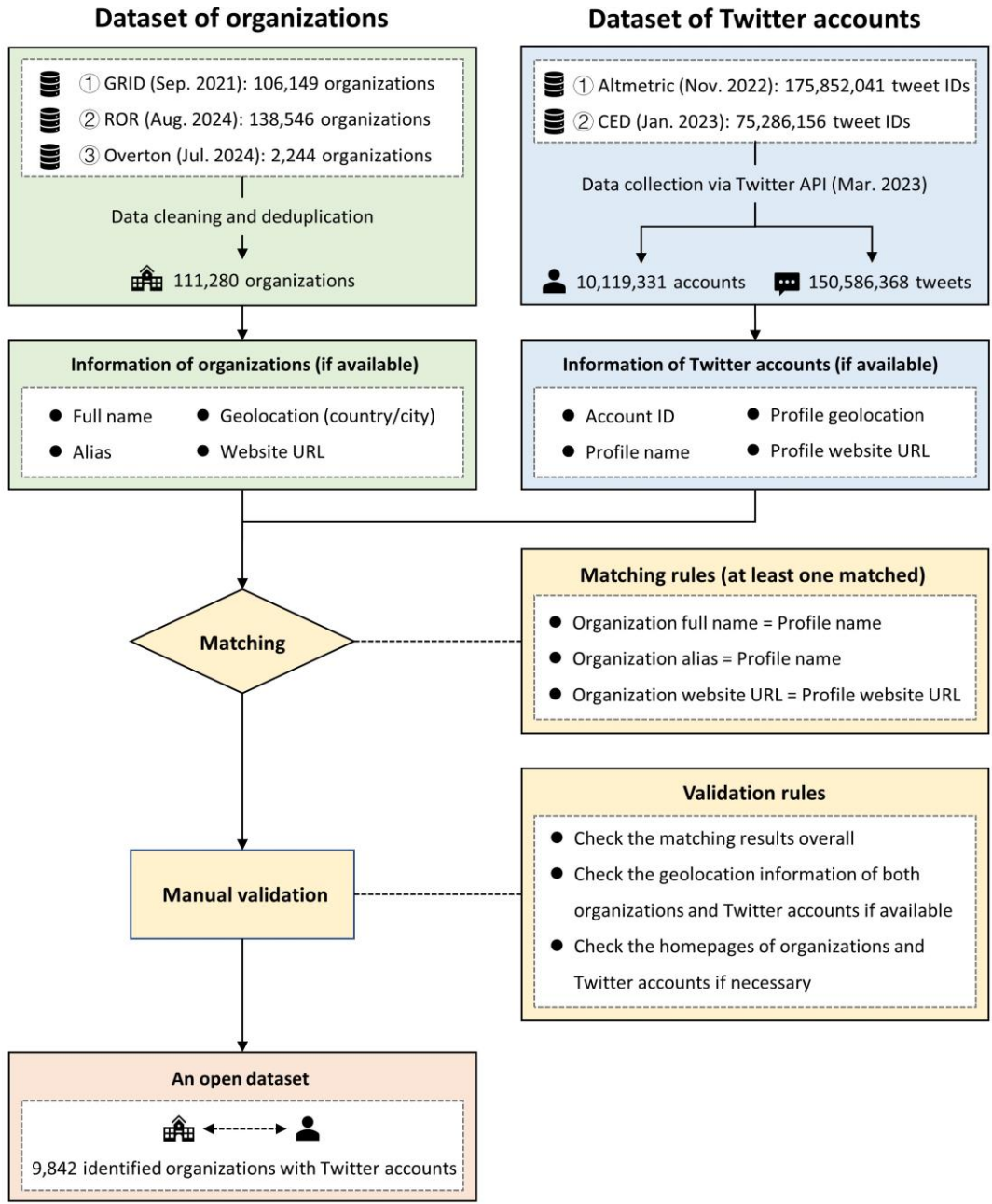

**Figure 1**. Workflow for identifying organisational accounts on Twitter. The snapshot versions of the databases used and their corresponding data volumes are indicated in the top frames, which represent the datasets of organisations and Twitter accounts, respectively.

*2.4. Categorization of organisations*

To analyze organisational activity on Twitter, we classified the identified organisations into seven main categories. This categorization was primarily derived from the ROR taxonomy[2]. For organisations indexed exclusively in Overton, we assigned categories according to the same scheme, using the organisation type information provided by Overton[3]. Table 1 presents the final classification system, including brief definitions, associated subcategories, and illustrative examples for each main category to clarify their composition.

**Table 1**. Categorization of organisations.

| Main category | Definition | Subcategories | Illustrative examples |
|---|---|---|---|
| Nonprofit | Independent organisations that operate for public, charitable, advocacy, or research-support purposes rather than for profit, often promoting research, education, or policy development. | Foundation; association; society; non-governmental organisation; think tank; charity | Age UK; Angel Foundation; Association of University Presses |
| Company | For-profit private-sector entities engaged in commercial, industrial, or financial activities that may support or apply scientific research. | Company; bank unaffiliated with government | IBM; Intel; Philips |
| Education | Academic institutions dedicated to teaching, learning, and research, encompassing higher education and professional training organisations. | University; college; school; polytechnic | ETH Zurich; University of Cambridge; Harvard University |
| Healthcare | Institutions that provide medical services, clinical care, and health-related research, often involved in producing or applying biomedical | Hospital; medical center | Cleveland Clinic; St. Vincent's Medical Center; Soroti Hospital |

---

[2] See more information about the organisation types of ROR from the documentation introducing the fields of ROR metadata: https://ror.readme.io/docs/fields.

[3] Among the 9,842 identified organisations with Twitter accounts, 9,531 had organisation type information available from ROR or GRID, allowing for direct classification according to the scheme presented in Table 1. For the remaining 311 organisations that were exclusively indexed in Overton, classification was based on Overton's own taxonomy. Overton categorizes its indexed organisations into three primary types: governments, inter-governmental organisations (IGOs), and think tanks. Following the scheme in Table 1, both governments and IGOs were assigned to the main category "Government", while think tanks were classified under "Nonprofit".

| | knowledge. | | |
|---|---|---|---|
| Facility | Specialized organisations or infrastructures focused on conducting, supporting, or enabling research, experimentation, observation, or data collection. | Lab; observatory; research institution; municipal facility | Vatican Observatory; Spiez Laboratory; Quadram Institute |
| Government | National or international public-sector bodies responsible for policymaking, regulation, governance, or public funding in areas including science, education, or societal affairs. | Government department; inter-governmental organisation; non-departmental public body | United Nations; Government of Spain; City of Sydney |
| Archive | Cultural and knowledge-preserving institutions that collect, curate, manage, and provide access to scientific, historical, or natural heritage resources. | Library; museum; archive; zoo; botanic garden | National Library of Australia; British Museum; National Zoological Park |

*2.5. Indicators*

Based on the detailed information collected for both scholarly tweets and Twitter accounts, we developed a set of indicators to measure the scholarly tweeting patterns of accounts. Following the framework proposed by Díaz-Faes et al. [17], the indicators were grouped into three dimensions: (1) *social media capital*, (2) *tweeting activity*, and (3) *engagement level*. Table 2 summarizes the indicators and their definitions.

**Table 2**. Indicators measuring the scholarly tweeting patterns of Twitter accounts.

| Dimension | Indicator | Definition |
|---|---|---|
| Social media capital | Followers | Number of users who follow an account. |
| | Followings | Number of users an account follows. |
| Tweeting activity | Annual tweets | Average number of tweets an account posts per year, calculated as the total number of tweets divided by the number of years between the account's creation year and the Twitter data collection year (2023). |
| | Annual scholarly | Average number of scholarly tweets an account posts per |

| | tweets | year, calculated as the total number of scholarly tweets divided by the number of years between the account's creation year and the Twitter data collection year (2023). Scholarly tweets are defined as those referencing scholarly publications, as recorded by Altmetric and CED. |
|---|---|---|
| | Originality rate | Proportion of original scholarly tweets among all scholarly tweets posted by an account. Higher values indicate a greater tendency to post original content rather than retweets. |
| | Scholarly focus rate | Proportion of scholarly tweets among all tweets posted by an account. Higher values indicate a stronger focus on disseminating scholarly publications. |
| Engagement level | Average likes | Average number of likes received per original scholarly tweet posted by an account. |
| | Average retweets | Average number of retweets received per original scholarly tweet posted by an account. |
| | Average quotes | Average number of quotes received per original scholarly tweet posted by an account. |
| | Average replies | Average number of replies received per original scholarly tweet posted by an account. |

Based on this classification:

(1) Social media capital: The number of *followers* and *followings* reflects an account's accumulated social media capital. Higher values indicate greater potential to attract attention and disseminate information to a broader audience.

(2) Tweeting activity: *Annual tweets* captures an account's overall activity on the platform since its creation, while the remaining three indicators focus specifically on scholarly communication. *Annual scholarly tweets* quantify the extent to which an account shares scholarly publications over time. *Originality rate* measures the extent to which an account posts original scholarly content rather than simply retweeting others' posts [54]. *Scholarly focus rate* reflects the proportion of an account's total tweets dedicated to scholarly publications, distinguishing research-oriented accounts from those sharing more general content.

(3) Engagement level: To assess the visibility and audience response to original scholarly tweets, we calculated *average likes*, *average retweets*, *average quotes*, and *average replies*. Higher values on these indicators suggest stronger visibility and interaction with scholarly content within the Twitter ecosystem.

*2.6. Analytical approaches*

To evaluate the scholarly tweeting behavior of organisational accounts within the broader Twitter landscape, we calculated relative measures of all indicators by considering the full set of 10,119,331 Twitter accounts in our database – that is, all users who had tweeted about scholarly publications at least once, as recorded by Altmetric and CED (see Section 2.2 for details on dataset construction). Because the absolute values of many indicators vary widely across accounts (see Table A1 in Appendix A), we computed the *percentile rank* for each indicator to enable meaningful comparisons.

The percentile rank positions each organisational account relative to the entire population of Twitter users engaged in scholarly communication. A higher percentile rank indicates stronger performance on a given indicator compared to the rest of the accounts. Percentile ranks for each indicator were calculated as follows:

- Compute the value of each indicator for every Twitter account, including both the identified organisational accounts and all other accounts (N = 10,119,331).

- Sort all accounts in ascending order based on the indicator values.

- Determine the percentile rank for each account, yielding values between 0 and 1.

A higher percentile rank thus reflects stronger performance relative to the full set of accounts, not only organisational ones. For example, to assess the relative activity level of an organisational account using the *annual tweets* indicator, we first calculate this value for all accounts in the dataset and then rank them in ascending order. If an organisational account achieves a percentile rank of 0.9, it has posted more tweets per year since its creation than 90% of all accounts, signifying a high level of tweeting activity.

In the subsequent analyses, we report only the percentile ranks of indicators achieved by the organisational accounts. It is important to note that, for each indicator examined, higher percentile ranks represent a *relative advantage* of organisational accounts compared with all other Twitter users engaged in scholarly communication, whereas lower percentile ranks indicate comparatively weaker performance.

## 3. Results

*3.1. An open dataset of organisational accounts on Twitter*

We identified 9,842 distinct organisational accounts that had tweeted about scholarly publications, thereby contributing to scholarly communication on Twitter. This dataset is openly available at https://doi.org/10.6084/m9.figshare.31321240 and is intended to facilitate future research on the presence and activities of organisational accounts on social media.

Figure 2 shows the distribution of these accounts across the seven organisational categories. Nonprofits represent the largest group (N = 3,353), followed by companies (N = 2,359) and educational organisations (N = 1,730). This distribution is consistent with earlier findings on health-related organisations engaged in Twitter discussions of health literacy [55], where nonprofits also emerged as the most active, followed by companies and educational organisations. These three categories thus appear as the most prominent organisational actors in scholarly communication on Twitter.

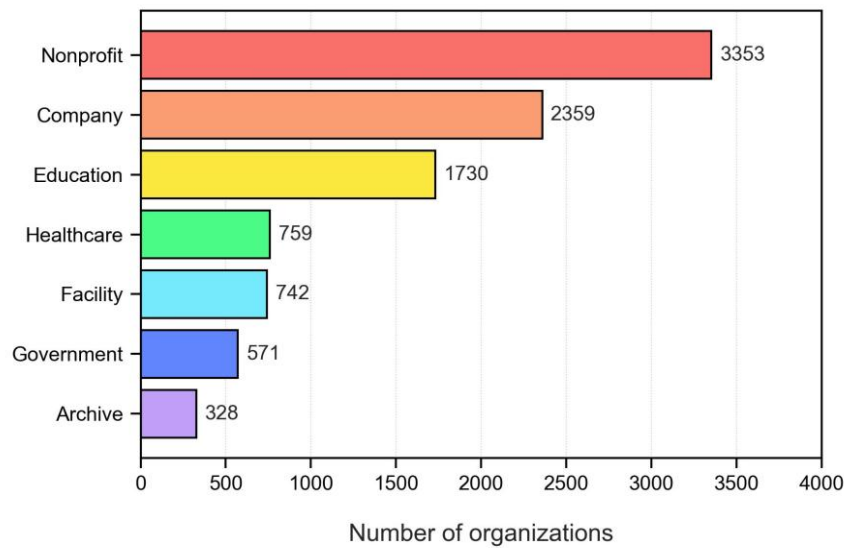

**Figure 2**. Distribution of the 9,842 identified organisational accounts across categories.

*3.2. Overall scholarly tweeting patterns of organisational accounts*

As shown in Figure 3, organisational accounts demonstrate a clear advantage in social media capital, as reflected by their relatively high percentile ranks for both *followers* and *followings* compared with all Twitter users engaged in scholarly communication. Most organisational accounts rank within the top 10% of users with respect to number of *followers*, highlighting their strong capacity to attract large audiences. Many also follow a comparatively larger number of users, although this behavior varies substantally across organisations, as indicated by the wide distribution of percentile ranks for *followings*.

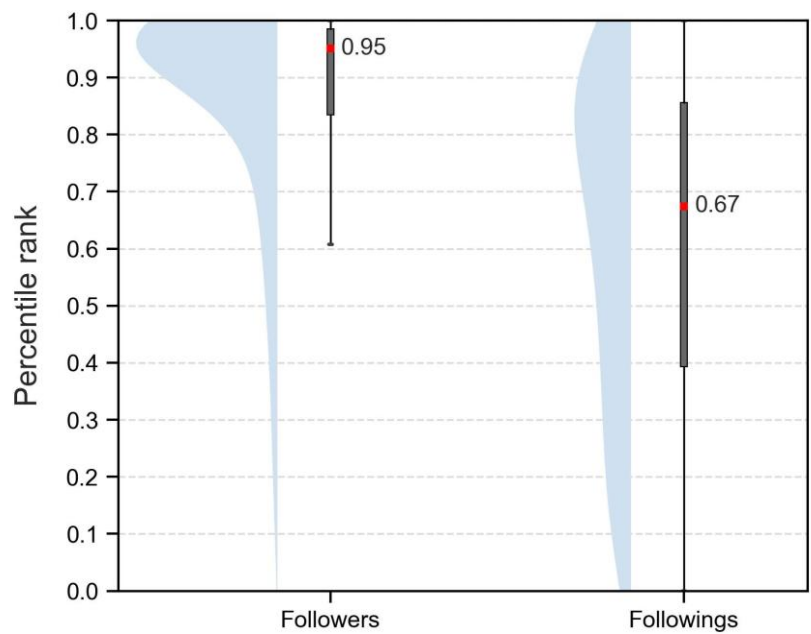

**Figure 3**. Percentile rank distributions of the 9,842 organisational accounts for social media capital, calculated relative to the full set of 10,119,331 Twitter accounts in the dataset. Median values are shown above each distribution.

Organisational accounts also exhibit distinct tweeting activity patterns (Figure 4). Although the distribution of *annual tweets* (median percentile rank = 0.50) indicates a moderate level of overall activity compared with other Twitter users, organisations are considerably more active in scholarly tweeting. This trend is particularly evident for *annual scholarly tweets*, *originality rate*, and *scholarly focus rate*, whose kernel density and boxplot distributions cluster above 0.6. In other words, organisational accounts tend to post more scholarly tweets per year, contribute a higher proportion of original content, and maintain a stronger focus on scholarly publications than the broader population of Twitter users engaged in scholarly communication.

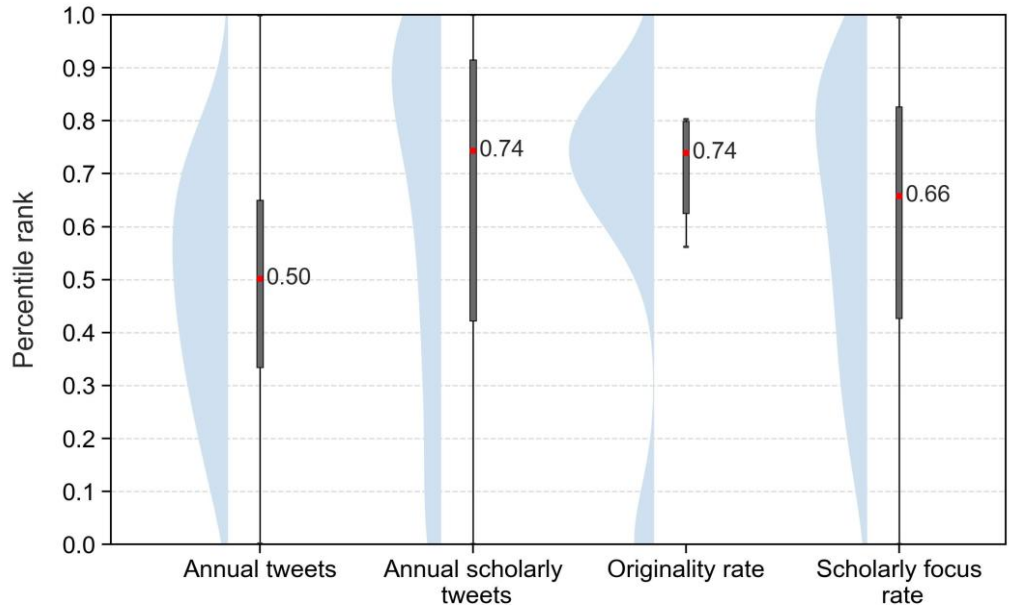

**Figure 4**. Percentile rank distributions of the 9,842 organisational accounts for tweeting activity, calculated relative to the full set of 10,119,331 Twitter accounts in the dataset. Median values are shown above each distribution.

In terms of engagement with original scholarly tweets (Figure 5), organisational accounts show contrasting patterns across the four metrics. Their scholarly tweets achieve high visibility through *average likes* and *average retweets*, with median percentile ranks of 0.79 and 0.86, respectively, placing nearly half of the organisations among the top 20% of all Twitter users. By contrast, for *average quotes* and *average replies* – forms of engagement that require deeper conversational interaction – only a small subset of organisations reach high levels (percentile ranks between 0.8 and 1), while the majority cluster near the lower end of the distribution. These results suggest that, although organisational tweets are effective in attracting attention, they are generally less successful in stimulating dialogue or sustained discussion.

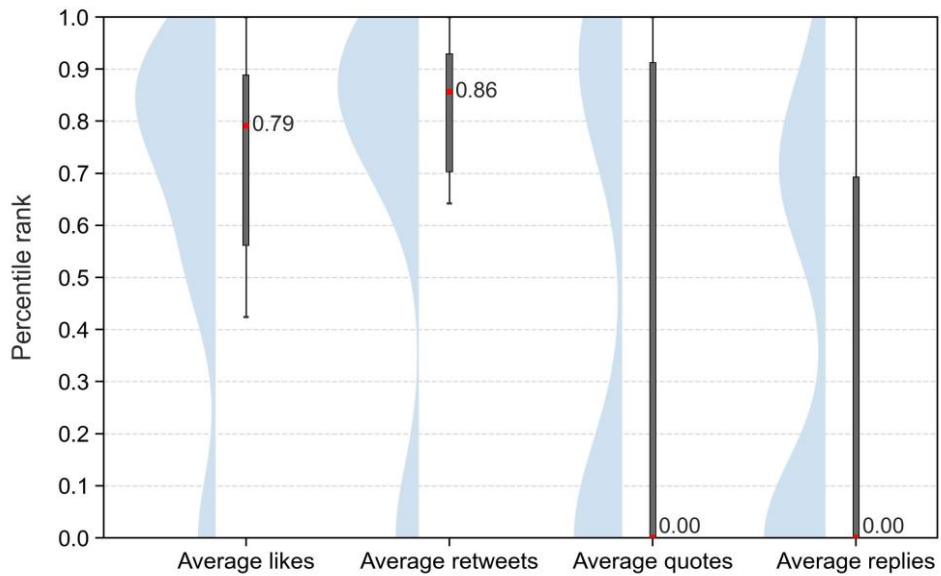

**Figure 5**. Percentile rank distributions of the 9,842 organisational accounts for engagement level, calculated relative to the full set of 10,119,331 Twitter accounts in the dataset. Median values are shown above each distribution.

*3.3. Scholarly tweeting patterns across different types of organisational accounts*

Across all categories, organisational accounts consistently possess greater social media capital than other Twitter users engaged in scholarly communication (Figure 6). This advantage is reflected in their relatively high percentile ranks for both *followers* and *followings*, particularly for the former. Nonetheless, differences emerge across categories. Governments, archives, nonprofits, and educational organisations generally attract larger follower bases than companies, healthcare institutions, and facilities. In addition, nonprofits and archives stand out for their higher levels of *followings*, suggesting stronger tendencies toward reciprocal connectivity and network building. The descriptive statistics by organisational category are consistent with the percentile rank results (see Table A2 in Appendix A): government organisations tend to accumulate the largest follower bases, while nonprofits and archives follow a greater number of other Twitter users, reinforcing their comparatively more interactive networking behavior.

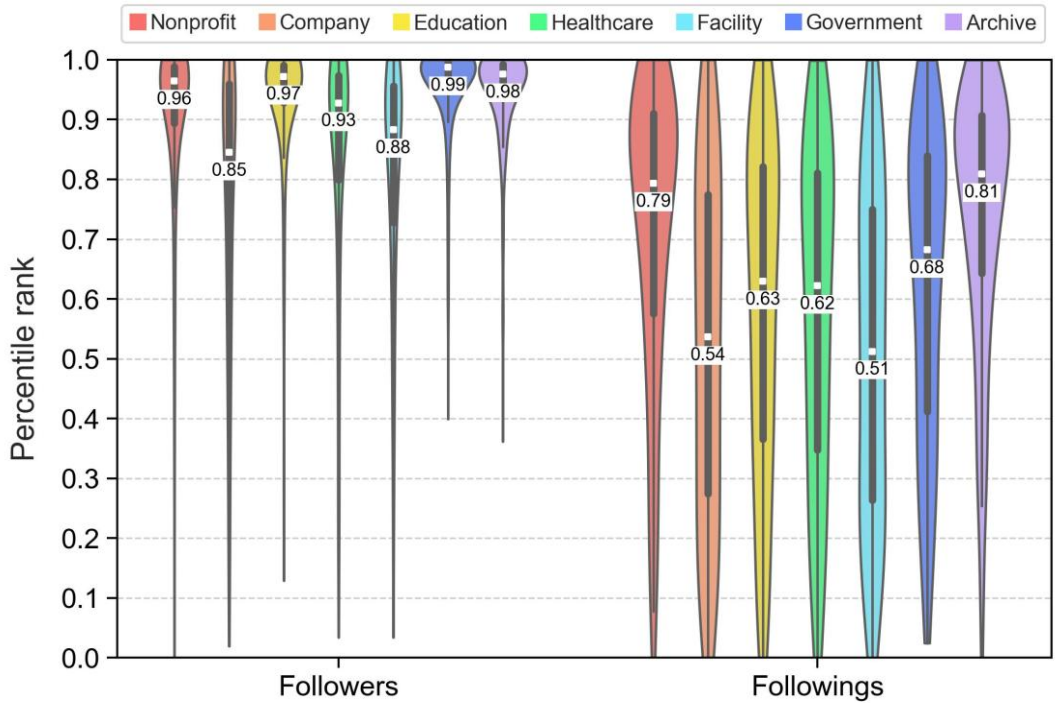

**Figure 6**. Percentile rank distributions of the seven categories of organisational accounts for social media capital, calculated relative to the full set of 10,119,331 Twitter accounts in the dataset. Median values are shown above each distribution.

Regarding tweeting activity, all seven organisational categories achieve relatively high levels of *originality rate*, but the other three indicators reveal more pronounced differences across groups (Figure 7). While most categories post a similar number of tweets per year as the overall population of Twitter users, companies and facilities exhibit comparatively lower levels of annual tweeting activity. However, facility accounts stand out by posting substantially more scholarly tweets per year, ranking highest among all categories. This strong emphasis on scholarly content also gives facilities the highest *scholarly focus rate*. In contrast, government and archive accounts display relatively lower *scholarly focus rates*, indicating that a greater share of their Twitter activity concerns topics beyond scholarly publications. These findings suggest that, although all organisational types participate in scholarly communication on Twitter, the degree of specialization and focus on scholarly content varies considerably by category. The percentile rank patterns are consistent with the descriptive statistics across organisational types (see Table A2 in Appendix A).

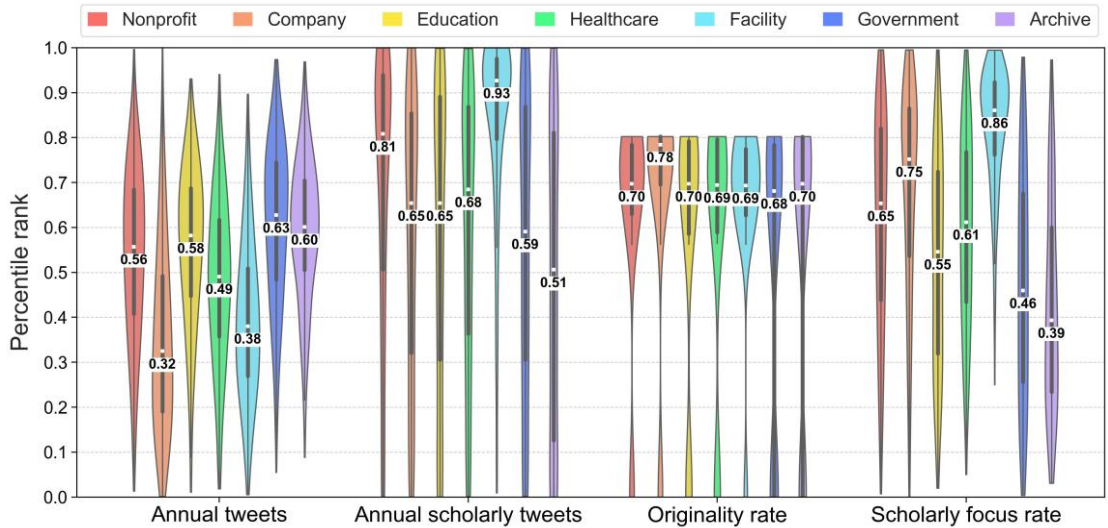

**Figure 7**. Percentile rank distributions of the seven categories of organisational accounts for tweeting activity, calculated relative to the full set of 10,119,331 Twitter accounts in the dataset. Median values are shown above each distribution.

Different categories of organisational accounts show engagement patterns with their scholarly tweets that broadly mirror the overall trends described above (Figure 8). In general, organisational accounts achieve high visibility through *average likes* and *average retweets*, with most categories ranking well above the median compared to other Twitter users engaged in scholarly communication. However, they tend to generate lower levels of conversational engagement, such as *average quotes* and *average replies*, suggesting a limited capacity to stimulate deeper interactions around scholarly content. Notable exceptions include government, facility, and nonprofit accounts. These categories perform above the median not only in likes and retweets but also in quotes and replies, indicating that, unlike other organisational types, they are comparatively more successful in fostering both broad visibility and interactive discussion on Twitter. The descriptive statistics in Table A2 (Appendix A) further support these patterns, particularly highlighting the strong engagement performance of government organisations, which show a distinct advantage across all four engagement metrics.

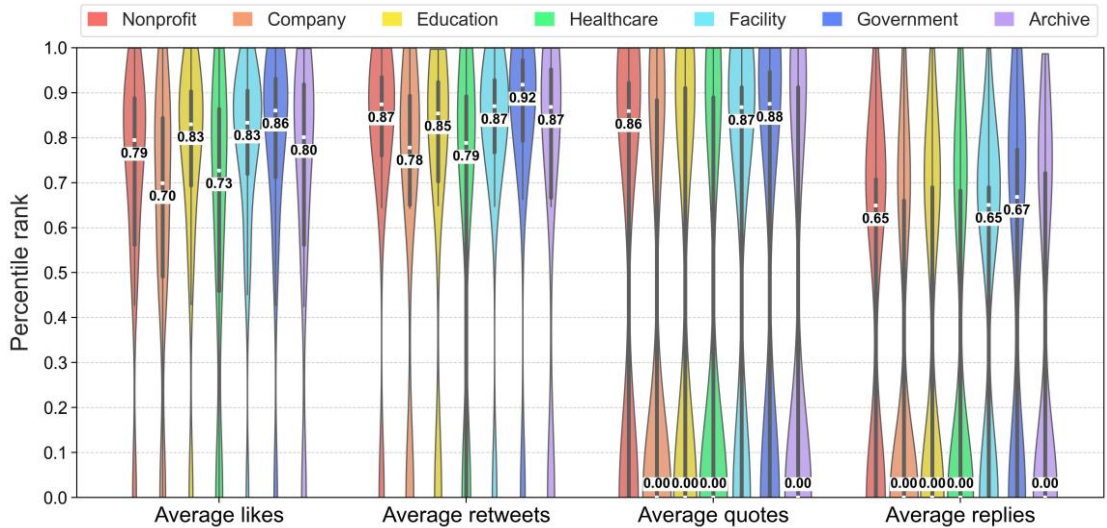

**Figure 8**. Percentile rank distributions of the seven categories of organisational accounts for engagement level, calculated relative to the full set of 10,119,331 Twitter accounts in the dataset. Median values are shown above each distribution.

## 4. Discussion

### 4.1. The presence and activity of organisations in scholarly communication

Due to the heterogeneity of social media platforms, user types, and motivations [7], it is widely recognized that characterizing users and their behaviors is essential for interpreting interactions surrounding scholarly works and understanding the nature of attention paid to them [17]. Against this backdrop, a substantial body of research has developed diverse methods to identify Twitter users engaged in scholarly communication, leading to the creation of open datasets focused on specific user groups, such as individual scholars [34], scholarly journals [38], and bots [48]. However, although previous work has noted that organisational accounts constitute a meaningful component of scholarly communication on Twitter [9, 27], no dedicated dataset of organisational accounts has been systematically developed. This gap has constrained more targeted empirical research on how organisations participate in and shape scholarly communication on social media.

Drawing on three organisational databases and two altmetric data sources, this study provides the first large-scale identification of research- and policy-related organisations involved in Twitter communication of scholarly publications. Unlike traditional altmetric studies that primarily examine how often scholarly works are mentioned on social media as a quantitative proxy for social attention or visibility, this study focuses on one key user community – organisational accounts – to characterize their presence, activity, and engagement in scholarly communication on Twitter. This perspective helps deepen our understanding of the nature of social media attention to scholarly works by foregrounding the types of actors who engage with it. Moreover, the resulting dataset of organisational accounts is openly available and can support future research on the presence and behavior of organisations not only in scholarly communication, but also in broader informational and public engagement contexts.

Out of 111,280 global organisations examined, we identified and validated 9,842 accounts (8.8%) that had tweeted about scholarly publications at least once, as recorded by Altmetric and CED. This indicates a relatively low level of organisational participation in Twitter-based scholarly communication, even among research- and policy-related organisations. This finding aligns with prior studies reporting that Twitter discussions of scholarly publications are largely dominated by individuals rather than organisations [6, 11, 13, 14]. For example, Htoo and Na [4] found that individuals accounted for 83% of users tweeting about psychological publications and contributed 77% of the scholarly tweets, compared to only 17% of organisational users, who contributed 23%. Taken together, these studies and our results reinforce the overall sparse presence of organisations in Twitter discussions of scholarly work.

By leveraging our dataset of organisational accounts alongside the full population of 10,119,331 Twitter accounts engaged in scholarly communication, we are able to assess the relative position of organisational actors in the broader dissemination of scholarly content – an aspect that has not previously been examined at this scale. Despite their limited presence, organisational accounts exhibit a stronger focus on scholarly publications than other users, as reflected in their higher annual scholarly tweet volumes and greater share of tweets referencing scholarly outputs. This pattern is partially attributable to our sampling frame: research- and policy-related organisations indexed in GRID, ROR, and Overton are more likely to use Twitter strategically to share, promote, or recommend research outputs [4]. The effect is particularly pronounced for research facilities, which show high levels of scholarly tweeting, but less so for government and archive accounts, which often tweet about administrative matters, events, or general news beyond scholarly discourse [43].

Organisational accounts also demonstrate substantial social media capital, attracting significantly larger follower bases than other Twitter users, which enhances their potential reach. This observation is consistent with earlier work showing that organisational accounts can reach audiences nearly three times larger than individual users [56]. In our dataset, this advantage translates into high visibility for scholarly tweets, as indicated by relatively high levels of likes and retweets. However, when examining conversational engagement measured through quotes and replies, most organisational accounts perform less strongly, suggesting that they largely adopt a broadcast-oriented communication style rather than one that actively encourages dialogue.

Several factors may explain this pattern. First, it reflects general engagement dynamics on Twitter: higher-effort engagement behaviors (e.g., quoting and replying) are substantially less common than lower-effort ones (e.g., liking and retweeting) [57]. Second, prior studies show that individual users are more likely to engage critically with scholarly publications [4], which tends to generate more conversational interaction. Third, because organisational accounts typically represent institutional voices, the individuals managing them may be constrained in how they interact publicly due to communication policies and reputational considerations, limiting opportunities for dialogic engagement.

However, exceptions do exist. Government, facility, and nonprofit accounts display higher levels of both visibility and conversational engagement. These groups not only receive more likes and retweets but also more quotes and replies than other organisational categories. While previous research has often characterized government and nonprofit communication on Twitter as predominantly one-way [58–61], our findings suggest that, within the context of scholarly communication, government accounts in particular are more successful in stimulating dialogue. This may be due to their large follower bases, public authority status, and the heightened motivation of audiences to engage with scholarly content disseminated by trusted or influential institutional actors. These results highlight a potentially important role for government organisations in facilitating public understanding and discussion of research.

*4.2. Implications of the research findings*

The findings of this study have several implications for how organisations use Twitter strategically and for how altmetrics should interpret organisational activity:

First, this study contributes an open dataset of organisational accounts engaged in scholarly communication, which can support future research investigating and monitoring organisational behaviors on Twitter. By linking organisational databases with altmetric data, the study also provides a reusable methodological framework for identifying and categorizing organisational accounts in large-scale datasets. This framework can be extended to other emerging platforms (e.g., Bluesky and Mastodon) and adapted to different organisational contexts, thereby enriching the methodological toolkit available to altmetric research.

Second, the findings show that organisations play an important role in scholarly communication on Twitter, as evidenced by the enhanced visibility of their tweets about scholarly publications. This visibility benefits from both their substantial follower bases and their relatively strong scholarly focus. Whereas traditional altmetrics have primarily emphasized publication-level or individual-level engagement, incorporating organisational activity will broaden the scope of analysis and provide a more comprehensive understanding of how diverse actors shape scholarly communication.

Third, given the growing interest in assessing the societal impact of research, organisations should be more systematically considered in research evaluation frameworks. Tweets by organisational accounts such as government agencies and certain nonprofits (e.g., think tanks) may serve as indicators of how research evidence is referenced in policy and decision-making contexts, particularly given the high levels of visibility and engagement these accounts achieve in relation to scholarly content in our study. As such, organisational accounts may act as intermediaries between science, policy, public debate, and professional practice, offering valuable signals for funding agencies and policymakers in the design of impact assessment models.

*4.3. Limitations and future work*

This study has several limitations that should be acknowledged:

Frist, we focused specifically on research- and policy-related organisations recorded in the GRID, ROR, and Overton databases. This scope inevitably excludes other types of organisations that may engage in scholarly communication on Twitter but are not covered by these registries. Moreover, the process of cross-walking across registries and matching them to Twitter profiles may introduce coverage and survivorship biases. Organisations with incomplete metadata, inconsistent or ambiguous names, recent structural changes (e.g., merger or renaming), or inactive/abandoned web presences are less likely to be successfully identified. Future research could expand the organisational sampling frame by integrating additional registries (e.g., national institutional directories) and by applying alternative account discovery strategies (e.g., follower network expansion, supervised or semi-supervised user classification) to improve coverage and reduce these biases.

Second, our analysis centered on organisational accounts that tweeted about scholarly publications, as recorded by Altmetric and CED. Although these are widely used altmetric data sources, they are not without limitations. Organisations with Twitter accounts that never shared publication-linked tweets were not included, which may lead to an underestimation of organisational engagement with research on social media. Furthermore, tweets that refer to research more implicitly (e.g., without linking to DOIs or URLs) are unlikely to be captured. This underscores the need for more comprehensive social media datasets that capture broader forms of scholarly discourse. Future research could therefore analyze a wider scope of organisational accounts, including those engaging with research in less direct ways.

Third, although our study provides large-scale quantitative evidence on organisational tweeting patterns and engagement dynamics, we did not examine the content and rhetorical strategies used in organisational tweets, such as hashtag practices, mention networks, or framing styles. Qualitative and linguistic analyses of tweet text and interaction structures would provide deeper insight into why some organisation types are more successful in generating conversational engagement, such as government, nonprofit, and facility accounts. Future work could build on the open dataset released in this study to conduct such fine-grained textual and interactional analyses.

Lastly, recent policy changes to Twitter following its rebranding as X – such as the introduction of a paid API and reduced content moderation – have prompted many academic users to migrate to alternative platforms like Bluesky and Mastodon [62]. Nevertheless, according to surveys by Altmetric, Twitter continues to host substantial discussions of research outputs [63, 64], which justifies the present study's focus. Moving forward, however, future research should increasingly examine newer platforms such as Bluesky and Mastodon to assess how organisations adapt their communication strategies in these emerging environments and whether the patterns identified here persist or change.

## 5. Conclusions

This study presents a large-scale identification and analysis of research- and policy-related organisations engaged in scholarly communication on Twitter. By integrating organisational registries (GRID, ROR, and Overton) with altmetric data sources (Altmetric and CED), we constructed an openly available dataset of 9,842 organisational accounts that had tweeted about scholarly publications.

Our findings reveal that these accounts demonstrate a strong focus on scholarly outputs, especially in the case of research facilities, and possess substantial social media capital, as reflected in their relatively large follower bases. This advantage enables organisational accounts to achieve high visibility for their scholarly tweets, as indicated by elevated levels of likes and retweets. However, most organisational accounts are less effective in generating conversational engagement, with comparatively low levels of quotes and replies. Notable exceptions include government organisations, which exhibit a greater capacity to promote both visibility and interactive discussion, suggesting their potential role in facilitating public engagement with research.

Beyond these empirical insights, the study contributes both an open dataset and a methodological framework that can support future investigations into the role of organisations in online scholarly communication. More broadly, the results underscore the importance of incorporating organisational activity into altmetric research, thereby expanding the analytical lens beyond publications and individuals to better capture how scholarly discourse is shaped by diverse institutional actors.

**Acknowledgments**

The authors thank Altmetric, Crossref Event Data, and Overton for providing the data for research purposes, and the anonymous reviewers for their valuable suggestions.

**Statements and Declarations**

Not applicable

**Declaration of conflicting interest**

The author(s) declared no potential conflicts of interest with respect to the research, authorship, and/or publication of this article.

**Funding statement**

The author(s) disclosed receipt of the following financial support for the research, authorship and/or publication of this article: This study was funded by the National Natural Science Foundation of China (No. 72304274). Er-Te Zheng was financially supported by the GTA scholarship from the School of Information, Journalism and Communication of the University of Sheffield.

**Appendix A**

For the ten indicators used to measure the scholarly tweeting patterns of organisational accounts, Table A1 reports the overall descriptive statistics, and Table A2 presents the statistics disaggregated by organisational category. Overall, the 9,842 organisational accounts exhibit substantial variation across most indicators, as reflected in the spread of the quartiles as well as the large standard deviations and skewness values (Table A1). These differences become even more pronounced when comparing across organisational types, with notable variation in both activity and engagement patterns (Table A2).

Table A1. Descriptive statistics for the ten indicators of the 9,842 identified organisational accounts

| Indicator | Mean | Min | Max | Q1 | Median | Q3 | IQR | Std. | Skewness |
|---|---|---|---|---|---|---|---|---|---|
| Followers | 49,098.69 | 0.00 | 65,688,604 | 1099.25 | 4033.00 | 14,205.25 | 13,106.00 | 837,425.54 | 58.75 |
| Followings | 1397.53 | 0.00 | 170,157 | 265.00 | 684.00 | 1497.50 | 1232.50 | 4159.73 | 23.24 |
| Annual tweets | 638.44 | 0.15 | 61,241 | 119.57 | 336.56 | 793.71 | 674.14 | 1226.39 | 19.89 |
| Annual scholarly tweets | 5.91 | 0.06 | 3452.73 | 0.18 | 0.60 | 2.62 | 2.44 | 46.85 | 48.35 |
| Originality rate | 0.54 | 0.00 | 1.00 | 0.22 | 0.52 | 0.92 | 0.70 | 0.36 | -0.12 |
| Scholarly focus rate | 0.02 | 0.00 | 0.85 | 0.00 | 0.00 | 0.01 | 0.01 | 0.05 | 6.65 |
| Average likes | 6.40 | 0.00 | 1772.08 | 1.00 | 2.60 | 5.67 | 4.67 | 36.00 | 33.27 |
| Average retweets | 2.69 | 0.00 | 532.67 | 0.33 | 1.15 | 2.65 | 2.32 | 11.06 | 28.28 |
| Average quotes | 0.27 | 0.00 | 148.00 | 0.00 | 0.00 | 0.25 | 0.25 | 1.95 | 57.66 |
| Average replies | 0.28 | 0.00 | 107.00 | 0.00 | 0.00 | 0.20 | 0.20 | 2.15 | 31.96 |

Note: Q1 denotes the 25th percentile, Q3 denotes the 75th percentile, and IQR refers to the interquartile range. Std. indicates standard deviation.

Table A2. Descriptive statistics for the ten indicators by organisational category

| Indicator | Type | Mean | Min | Max | Q1 | Median | Q3 | IQR | Std. | Skewness |
|---|---|---|---|---|---|---|---|---|---|---|
| Followers | Nonprofit | 32,078.83 | 0.00 | 5,086,941.00 | 1819.00 | 5697.00 | 17,267.00 | 15,448.00 | 190,511.84 | 18.90 |
| | Company | 84,684.08 | 2.00 | **65,688,604.00** | 339.50 | 1191.00 | 4914.50 | 4575.00 | **1,630,336.93** | **32.58** |
| | Education | 26,215.91 | 22.00 | 1,573,124.00 | 2759.75 | 7343.00 | 21,702.75 | 18,943.00 | 74,626.48 | 10.44 |
| | Healthcare | 11,868.62 | 4.00 | 1,874,657.00 | 863.00 | 2736.00 | 7763.00 | 6900.00 | 77,459.45 | 20.00 |
| | Facility | 17,208.78 | 4.00 | 7,704,373.00 | 567.50 | 1638.50 | 4566.00 | 3998.50 | 283,629.99 | 26.95 |
| | Government | **156,266.52** | **126.00** | 16,277,981.00 | **4605.50** | **16,134.00** | **67,072.50** | **62,467.00** | 848,587.46 | 14.68 |
| | Archive | 59,574.17 | 105.00 | 2,292,711.00 | 3094.00 | 8488.50 | 26,998.75 | 23,904.75 | 211,166.41 | 7.48 |
| Followings | Nonprofit | **1960.77** | 0.00 | **170,157.00** | 488.00 | 1084.00 | **2155.00** | **1667.00** | 5257.78 | 19.11 |
| | Company | 1098.06 | 0.00 | 148,133.00 | 166.00 | 429.00 | 998.00 | 832.00 | 4865.13 | **24.07** |
| | Education | 1103.18 | 0.00 | 53,771.00 | 239.25 | 584.50 | 1241.75 | 1002.50 | 2348.65 | 13.83 |
| | Healthcare | 1011.44 | 0.00 | 21,809.00 | 224.00 | 570.00 | 1176.00 | 952.00 | 1529.81 | 5.72 |
| | Facility | 813.42 | 0.00 | 38,225.00 | 158.00 | 396.50 | 906.00 | 748.00 | 1853.15 | 12.99 |
| | Government | 1217.29 | **7.00** | 30,839.00 | 283.00 | 703.00 | 1362.50 | 1079.50 | 2040.61 | 7.44 |
| | Archive | 1874.81 | 0.00 | 46,295.00 | **612.25** | **1164.00** | 2105.50 | 1493.25 | 3359.86 | 8.88 |
| Annual tweets | Nonprofit | 768.94 | 1.00 | 35,513.85 | 192.86 | 464.36 | 977.00 | 784.14 | 1256.88 | 13.91 |
|  | Company | 382.38 | 0.15 | **61,241.00** | 38.00 | 112.57 | 319.46 | 281.46 | **1640.78** | **24.94** |
|  | Education | 732.59 | 0.75 | 6739.00 | 245.93 | 536.45 | 996.73 | 750.80 | 684.53 | 1.98 |
|  | Healthcare | 503.83 | 1.33 | 7672.23 | 139.76 | 317.07 | 655.88 | 516.12 | 613.56 | 4.08 |
|  | Facility | 313.20 | 0.40 | 4597.71 | 75.05 | 162.49 | 352.39 | 277.34 | 462.21 | 4.07 |
|  | Government | **1098.28** | 5.20 | 13,775.53 | 304.01 | **697.80** | **1426.37** | **1122.36** | 1362.89 | 4.02 |
|  | Archive | 896.10 | **10.38** | 12,304.67 | **344.73** | 597.73 | 1103.89 | 759.16 | 1082.46 | 5.16 |
| Annual scholarly tweets | Nonprofit | 8.72 | 0.06 | **3452.73** | 0.25 | 0.92 | 4.00 | 3.75 | **68.83** | **39.07** |
|  | Company | 4.53 | 0.06 | 1719.60 | 0.14 | 0.40 | 1.33 | 1.19 | 44.33 | 28.49 |
|  | Education | 2.83 | 0.06 | 140.47 | 0.13 | 0.40 | 2.00 | 1.87 | 7.91 | 7.45 |
|  | Healthcare | 2.87 | 0.06 | 100.85 | 0.15 | 0.46 | 1.52 | 1.37 | 8.95 | 6.19 |
|  | Facility | **11.32** | **0.07** | 576.87 | **0.82** | **3.14** | **10.60** | **9.78** | 29.18 | 11.52 |
|  | Government | 3.52 | 0.06 | 258.50 | 0.13 | 0.33 | 1.50 | 1.37 | 15.41 | 11.24 |
|  | Archive | 2.22 | 0.06 | 238.17 | 0.08 | 0.24 | 1.00 | 0.92 | 14.01 | 15.15 |
| Originality rate | Nonprofit | 0.51 | 0.00 | 1.00 | 0.25 | 0.50 | 0.80 | 0.55 | 0.34 | -0.02 |
|  | Company | **0.68** | 0.00 | 1.00 | **0.47** | **0.80** | **1.00** | 0.53 | 0.36 | **-0.81** |
|  | Education | 0.48 | 0.00 | 1.00 | 0.10 | 0.50 | 0.86 | 0.76 | 0.37 | 0.11 |
|  | Healthcare | 0.48 | 0.00 | 1.00 | 0.11 | 0.46 | 0.89 | 0.78 | 0.38 | 0.17 |
|  | Facility | 0.48 | 0.00 | 1.00 | 0.23 | 0.45 | 0.74 | 0.51 | 0.32 | 0.20 |
|  | Government | 0.43 | 0.00 | 1.00 | 0.00 | 0.40 | 0.80 | 0.80 | 0.39 | 0.27 |
|  | Archive | 0.49 | 0.00 | 1.00 | 0.00 | 0.50 | 0.98 | **0.98** | 0.39 | 0.03 |
| Scholarly focus rate | Nonprofit | 0.02 | 0.00 | **0.85** | 0.00 | 0.00 | 0.01 | 0.01 | 0.05 | 7.22 |
| | Company | 0.03 | 0.00 | 0.81 | 0.00 | 0.01 | 0.02 | 0.02 | 0.06 | 5.13 |
| | Education | 0.01 | 0.00 | 0.83 | 0.00 | 0.00 | 0.00 | 0.00 | 0.05 | **11.33** |
| | Healthcare | 0.01 | 0.00 | 0.77 | 0.00 | 0.00 | 0.01 | 0.01 | 0.04 | 11.22 |
| | Facility | **0.05** | 0.00 | 0.74 | **0.01** | **0.02** | **0.06** | **0.06** | **0.08** | 3.49 |
| | Government | 0.01 | 0.00 | 0.27 | 0.00 | 0.00 | 0.00 | 0.00 | 0.02 | 8.02 |
| | Archive | 0.00 | 0.00 | 0.22 | 0.00 | 0.00 | 0.00 | 0.00 | 0.02 | 7.54 |
| Average likes | Nonprofit | 5.40 | 0.00 | 442.50 | 1.00 | 2.74 | 5.58 | 4.58 | 13.66 | 16.51 |
| | Company | 6.69 | 0.00 | **1772.08** | 0.50 | 1.57 | 4.00 | 3.50 | 52.07 | 25.42 |
| | Education | 6.03 | 0.00 | 446.50 | 1.46 | 3.33 | 6.43 | 4.97 | 15.27 | 19.17 |
| | Healthcare | 4.85 | 0.00 | 471.81 | 0.26 | 2.00 | 4.50 | 4.24 | 20.93 | 18.99 |
| | Facility | 7.58 | 0.00 | 1550.00 | **1.78** | 3.64 | 6.56 | 4.78 | 59.46 | **25.54** |
| | Government | **13.06** | 0.00 | 1119.00 | 1.67 | **4.19** | **9.00** | **7.33** | 61.34 | 15.60 |
| | Archive | 7.79 | 0.00 | 173.50 | 1.00 | 3.00 | 7.54 | 6.54 | 18.06 | 6.14 |
| Average retweets | Nonprofit | 2.92 | 0.00 | 176.17 | 0.58 | 1.50 | 3.08 | 2.50 | 7.02 | 13.86 |
| | Company | 2.36 | 0.00 | **532.67** | 0.08 | 0.75 | 2.00 | 1.92 | 14.58 | **26.68** |
| | Education | 1.92 | 0.00 | 26.00 | 0.33 | 1.10 | 2.48 | 2.15 | 2.75 | 3.78 |
| | Healthcare | 2.13 | 0.00 | 300.84 | 0.00 | 0.92 | 1.91 | 1.91 | 12.60 | 22.34 |
| | Facility | 2.35 | 0.00 | 184.00 | 0.65 | 1.43 | 2.66 | 2.01 | 7.56 | 20.83 |
| | Government | **6.50** | 0.00 | 455.00 | **1.00** | 2.15 | 5.41 | 4.41 | 25.77 | 14.26 |
|  | Archive | 3.25 | 0.00 | 116.00 | 0.15 | 1.39 | 4.00 | 3.85 | 8.33 | 10.72 |
| Average quotes | Nonprofit | 0.25 | 0.00 | 17.40 | 0.00 | 0.06 | 0.30 | 0.30 | 0.67 | 14.75 |
|  | Company | 0.25 | 0.00 | 60.08 | 0.00 | 0.00 | 0.13 | 0.13 | 1.70 | **24.44** |
|  | Education | 0.21 | 0.00 | 10.50 | 0.00 | 0.00 | 0.24 | 0.24 | 0.54 | 10.23 |
|  | Healthcare | 0.22 | 0.00 | 28.16 | 0.00 | 0.00 | 0.15 | 0.15 | 1.22 | 20.29 |
|  | Facility | 0.21 | 0.00 | 8.79 | 0.00 | 0.09 | 0.25 | 0.25 | 0.47 | 10.17 |
|  | Government | **0.88** | 0.00 | **148.00** | 0.00 | **0.11** | **0.50** | **0.50** | 7.54 | 19.06 |
|  | Archive | 0.26 | 0.00 | 11.00 | 0.00 | 0.00 | 0.25 | 0.25 | 0.82 | 10.04 |
| Average replies | Nonprofit | 0.25 | 0.00 | 60.50 | 0.00 | 0.05 | 0.25 | 0.25 | 1.34 | **33.98** |
|  | Company | 0.29 | 0.00 | 64.33 | 0.00 | 0.00 | 0.09 | 0.09 | 2.32 | 19.33 |
|  | Education | 0.20 | 0.00 | 15.00 | 0.00 | 0.00 | 0.18 | 0.18 | 0.69 | 12.19 |
|  | Healthcare | 0.26 | 0.00 | 36.00 | 0.00 | 0.00 | 0.17 | 0.17 | 1.55 | 20.51 |
|  | Facility | 0.16 | 0.00 | 14.00 | 0.00 | 0.05 | 0.18 | 0.18 | 0.59 | 19.16 |
|  | Government | **1.04** | 0.00 | **107.00** | 0.00 | **0.12** | **0.50** | **0.50** | 7.02 | 13.31 |
|  | Archive | 0.21 | 0.00 | 2.33 | 0.00 | 0.00 | 0.26 | 0.26 | 0.38 | 2.58 |

Note: Q1 denotes the 25th percentile, Q3 denotes the 75th percentile, and IQR refers to the interquartile range. Std. indicates standard deviation. For each indicator, the largest absolute value across organisation types is highlighted in bold, where applicable.